\documentclass[lettersize,journal]{IEEEtran}
\usepackage{amsmath,amsfonts}
\usepackage{algorithmic}
\usepackage{array}
\usepackage{subfig}
\usepackage{textcomp}
\usepackage{stfloats}
\usepackage{url}
\usepackage{verbatim}
\usepackage{graphicx}
\def\BibTeX{{\rm B\kern-.05em{\sc i\kern-.025em b}\kern-.08em
    T\kern-.1667em\lower.7ex\hbox{E}\kern-.125emX}}
\usepackage{balance}
\begin{document}
\title{Semantic Sensing and Communications for Ultimate Extended Reality}
\author{Bowen Zhang, Zhijin Qin, Yiyu Guo, Geoffrey Ye Li
}


\maketitle

\begin{abstract}
As a key technology in metaversa, wireless ultimate extended reality (XR) has attracted extensive attentions from both industry and academia. However, the stringent latency and ultra-high data rates requirements have hindered the development of wireless ultimate XR. Instead of transmitting the original source data bit-by-bit, semantic communications focus on the successful delivery of semantic information contained in the source, which have shown great potentials in reducing the data traffic of wireless systems. Inspired by semantic communications, this article develops a joint semantic sensing, rendering, and communication framework for wireless ultimate XR. In particular, semantic sensing is used to improve the sensing efficiency by exploring the spatial-temporal distributions of semantic information. Semantic rendering is designed to reduce the costs on semantically-redundant pixels. Next, semantic communications are adopted for high data transmission efficiency in wireless ultimate XR. Then, two case studies are provided to demonstrate the effectiveness of the proposed framework. Finally, potential research directions are identified to boost the development of semantic-aware wireless ultimate XR.     
\end{abstract}


\section{Introduction}
Extended reality (XR) is changing the way that humans interact with the virtual world and bringing the revolution to immersive gaming, videoconferencing, and remote shopping. XR technologies can be categorized into augmented reality (AR), mixed reality (MR), and virtual reality (VR), dependent on the portions of virtual contents to be rendered and displayed~\cite{akyildiz2022wireless}. The rendered virtual contents are usually displayed by XR head-mounted devices (HMDs). As the users have to wear these devices sometimes for a long time, XR HMDs have strict restrictions on their weight, energy consumption, and heat dissipation. To meet the restrictions, HMDs have to offload most of the storage and computing tasks to remote servers, which requires ultra low latency connections between HMDs and remote servers.

Recently, high-quality wireless ultimate XR has been developed~\cite{akyildiz2022wireless,morin2022toward}. As HMDs are connected to remote servers through wireless networks, users can move freely with better quality-of-experience (QoE). However, wireless ultimate XR still faces some critical challenges. First, ultimate XR requires the motion-to-photo latency to be less than $10$ ms~\cite{morin2022toward}, which means the corresponding changes caused by users' controlling commands or the eye/head/foot movements should be presented to the users within $10$ ms. However, latency in nearly all components in a wireless XR system, including sensing, rendering, and communication parts makes the ultra-low latency difficult to achieve when generating and transmitting virtual contents with high resolutions. Secondly, the data transmission rate is expected to exceed tens of Gbps in ultimate XR~\cite{morin2022toward}, which is far above the achievable capacity of the existing wireless networks with peak from 0.1 Gbps to 2.0 Gbps~\cite{akyildiz2022wireless}. Last but not least, the high power consumption caused by data processing and transmission also limits the lifetime of HMDs. 
Therefore, a new communication network architecture for supporting wireless ultimate XR is more than desired.

To overcome the ultra-high data rate challenges in wireless ultimate XR, semantic communication~\cite{qin2021semantic} is considered as a promising solution. Different from conventional communications, semantic communications only transmit semantic information relevant to the task at the receiver, significantly reducing the volume of data traffic. 
Recent studies have shown that semantic communication systems can greatly improve the transmission efficiency in serving various applications, such as wireless image retrieval~\cite{jankowski2020wireless}, video conferencing~\cite{jiang2022wireless}, and object detection~\cite{qin2021semantic}. In wireless ultimate XR, large amounts of sensed data and rendered results will be transmitted between XR HMDs and the remote server for computing tasks and view synthesis tasks. By exploring and extracting the task-relevant semantic information from the sensed data and rendered results, the size of data to be transmitted can be greatly reduced.


Despite the attractive performance of semantic communications, they are insufficient for satisfying the latency and power consumption restrictions in wireless ultimate XR systems. In ultimate XR systems, high-resolution images/videos are frequently captured or rendered. Generating and processing these high-resolution sensed or rendered images/videos introduce significant latency and heavy power consumption to the system. For example, 360$^{\circ}$ stereo images are required for precise object insertion in AR/MR applications~\cite{li2021lighting}. Color and depth images in $2$K resolution are captured for 3D reconstruction~\cite{morin2022toward}. Full-view $24$K video need to be rendered to provide VR users immersive perceptual experience~\cite{akyildiz2022wireless}. Therefore, it is imperative to design fast and energy-efficient sensing and rendering techniques to put wireless ultimate XR into practice.

In this article, we propose a framework, including semantic sensing, semantic rendering, and semantic communications, to support wireless ultimate XR. The proposed framework can alleviate the sensing costs at HMDs, fasten the rendering process at the server, and reduce the data traffic between HMDs and the server. 

The rest of this article is organized as follows. Section~\ref{section II} discusses the general architecture and challenges of wireless ultimate XR. In Section~\ref{section III}, we demonstrate the proposed framework with semantic sensing, semantic rendering, and semantic communications, followed by user case studies in Section~\ref{section IV}. Section~\ref{IIII} concludes this article with potential research directions.

\section{Wireless Ultimate XR Architectures}
\label{section II}
\begin{figure*}[!t]
\centering
\includegraphics[width=0.8\textwidth]{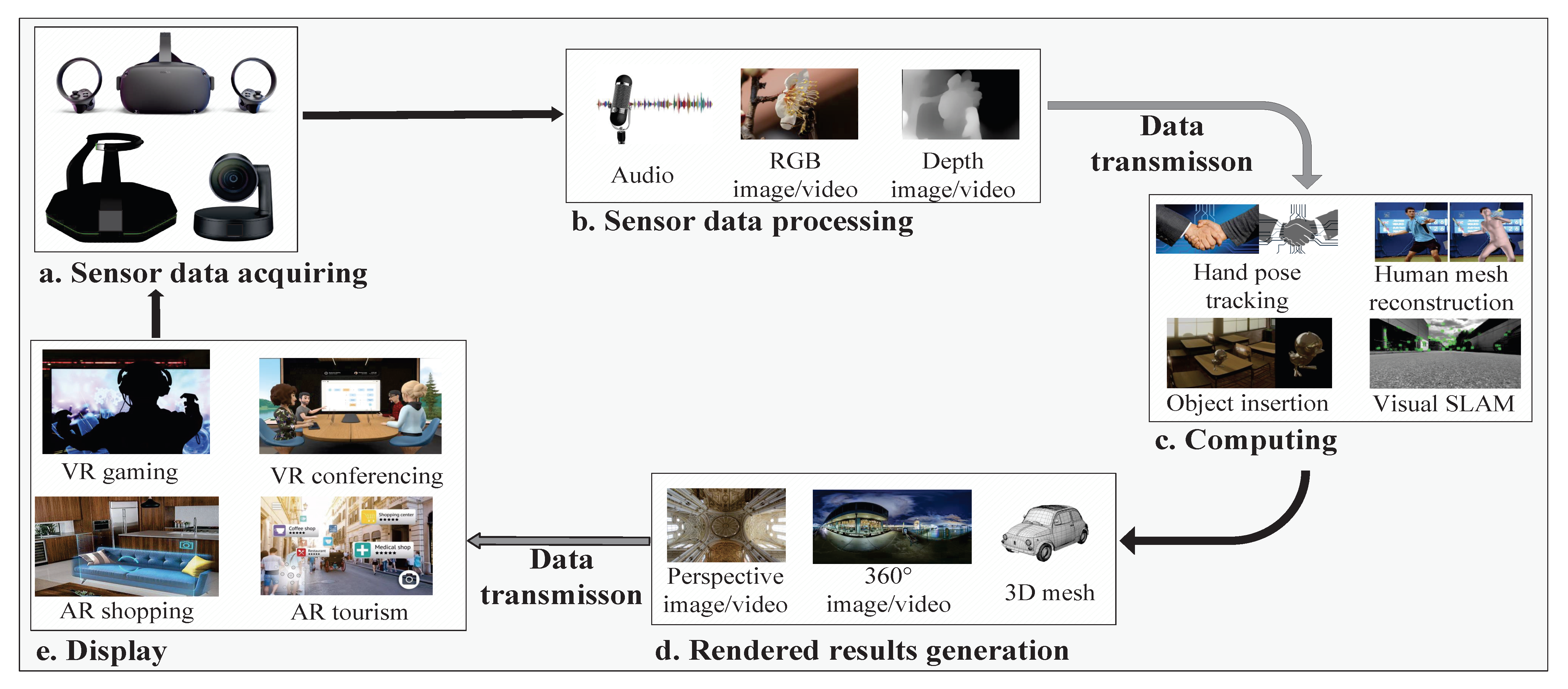}
\caption{A general wireless ultimate XR architecture. }
\label{fig1}
\end{figure*}

The procedures for wireless XR applications usually consists of sensing, transmission, computing, rendering, and displaying, as shown in Fig.~\ref{fig1}. To facilitate the introduction of our semantic based framework, we will provide a brief description on the wireless ultimate XR.

\subsection{Sensor Data Acquisition}
To map objects and scenes in user’s environments into the virtual display, some sensors are required at the user side, including microphones, cameras, and depth sensors. The amount of sensed data generated by all types of sensors can be quite large. For example, in ultimate VR video conferencing, high-resolution images of the users should be captured, helping the remote servers to add body, facial expressions, and dressing details to the users' virtual images. In ultimate AR shopping, high-resolution color and depth images of indoor environment are required by the servers, so that the servers can estimate the three-dimensional (3D) geometric information of the indoor environment, and display the virtual products correctly with reasonable shadows and occlusion~\cite{morin2022toward}.    

Generating and processing a large amount of sensor data is energy- and time-consuming for HMDs, especially when these sensory data needs to be further compressed and transmitted. Therefore, it is imperative to design a faster and power-efficient sensing system.    

\subsection{Sensed Data Processing and Transmission}
After collecting the sensed raw data, data aggregation, coding, and compression will be performed at HMDs, before they are uploaded to the remote servers for post-processing. However, the stringent latency restriction of ultimate XR applications poses an extremely high requirement on the wireless transmission rates. For example, if the sensed data is an image with $2$K resolution and is encoded by Better Portable Graphics (BPG) with $0.8$ bit per pixel (bpp), the compressed sensed data size will be of $1.6$ Mb. Considering the sensed image will affect the virtual contents, this transmission follows $10$ ms motion-to-photo requirement. If we further consider the time spent on rendering, downlink transmission, and displaying, the time budget for uplink transmission will be less than $1.25$ ms, which requires the data rate of about $1.3$ Gbps. In the ultimate XR, 360$^{\circ}$ stereo images and depth images will be used, which results in a large volume of sensed data and requires higher data transmission rate. 

\subsection{Computing}
Once the sensed data is received by the remote servers, a series of computing tasks will be conducted to extract useful information from these sensor data, which are used to construct or update various 3D models of objects and scenes used in virtual worlds. For example, hand pose estimation and tracking are widely used in VR gaming and conferencing. In this case, hand-related information will be extracted from the sensed data, and then used to construct a 3D hand model. As only task-related information in sensory data will be utilized at the server, it inspires us to reduce the transmitted data size in uplink connections through semantic communications.

\subsection{Rendered Results Generation and Transmission}
After obtaining the 3D models or scenes through computation, rendering will generate photo-realistic viewpoint-dependent perspective images or 360$^{\circ}$ panoramic images for users to perceive 3D scenes. Viewpoint-dependent images only contain the visual contents at a specific viewing direction, which, however, are not suitable for VR applications where users' heads move frequently. By contrast, rendering 360$^{\circ}$ images can deal with this problem with the cost of three times data rate. In ultimate AR/MR applications, rendering means adding light or shadow effects to 3D models to make it natural in real-world environment based on the computing results. 

Similar to the sensed data transmission, transmitting high-resolution rendered images or 3D models of objects and scenes under a stringent latency requirement is challenging. As stated in~\cite{akyildiz2022wireless}, full-view $24$K images need to be transmitted within about $2$-$4$ ms for downlink transmission in ultimate VR applications, which requires the data rate of more than $50$ Gbps. A medium-size 3D model stored in Polygon File Format will exceed $110$ Mb, resulting in the $30$ Gbps data rate requirement in AR applications. 

\subsection{Displaying}
The last step is to display these rendered virtual contents on HMDs for users. Sometimes, local processing is also required to ensure the smooth displaying. For instance, 3D world locking is required when displaying 3D models of virtual objects in AR/MR applications, so that the placement of these virtual objects would not change with users' movements. Meanwhile, view synthesis techniques are also required when users' fields of view (FoV) are different from what are expected in the rendering process. Therefore, displaying will also cost certain time and computational resources. 


\section{Semantic sensing, rendering, and communications}
\label{section III}
From section \ref{section III}, it is necessary to introduce semantic framework to address various challenges on implementing wireless ultimate XR. In this section, we will introduce the proposed semantic sensing, rendering, and communication architecture for wireless ultimate XR. An overall architecture is shown in Fig~\ref{fig2}.
\subsection{Semantic Sensing}
Processing a large amount of sensed data is challenging when HMDs have limited storage resources, energy capacity, and computing performances. Therefore, reducing the amount of data acquired at the sensor level should be considered. Recently, image or video compressed sensing has drawn considerable attention as an effective framework to alleviate the computational and sampling costs of sensors~\cite{martel2020neural,yang2018admm}. Different from the traditional sensing-then-compression pipeline, these works propose to conduct image or video sensing and compression simultaneously using compressed sensing, leading to a smaller sensory data size. 
The existing studies motivate us to rethink the sensing-then-task-execution framework, where plenty of task-irrelevant information will be also sensed, but discarded during the task-execution process. 

\subsubsection{Design Principles}
To improve the sampling efficiency, we propose the idea of semantic sensing, where only task-related signals will be extracted.
Semantic sensing can be divided into two categories: prior-based sensing and feedback-based sensing. In prior-based sensing, the statistical distribution of task-related information is summarised from a task-related dataset, and then used to design a spatial-temporal 3D binary mask. Subsequently, the 3D binary mask is compiled into a programmable sensor and used to selectively switch off some sensing elements in spatial-temporal dimensions, where the samples are semantically-redundant for a specific task, as shown in Fig.~\ref{fig2}. We now illustrate the idea of semantic sensing in more details by taking hand pose estimation and tracking as an example. Usually, the actions of user hands follow some fixed patterns in VR driving or shooting games. 
Studying these patterns can help us summarize the activity area and moving speed of user hands. Based on these prior information, we can selectively turn off some camera sensor elements in the spatial-temporal dimensions to reduce the sampling rates without affecting task accuracy. 

For the feedback-based sensing, the historical sensed results are used to infer the distribution of task-related samples, and guide the sensing process at the current timeslot. Specifically, when the locations of the interesting objects are unknown in captured images/videos, we can first collect all samples. After applying an object detection algorithm at the server side, the bounding box of the interested objects will be fed back to the sensors, which can then be used to reduce the number of samples to be taken.

\subsubsection{Architectural Details}
To implement semantic sensing, a mask generation network will be deployed at the HMDs, which is used to  
generate task-specific 3D binary masks by taking task tokens as inputs. For feedback-based sensing, this neural network will also take the estimated areas of interesting objects and historical task performance as inputs. The generated mask will dynamically adjust the sensing policy for each task, and then the sensed data will be fed into a task execution network. The task accuracy will be used to guide the training process of the mask generation network with the goal of maximizing the sparsity level of the binary mask under a task accuracy constraint. The semantic sensing module should also be jointly designed with the semantic communication modules, so that the latter can handle the changes of sensing policy.

\subsection{Semantic Rendering}
\begin{figure*}[t]
\centering
\includegraphics[width=.9\textwidth]{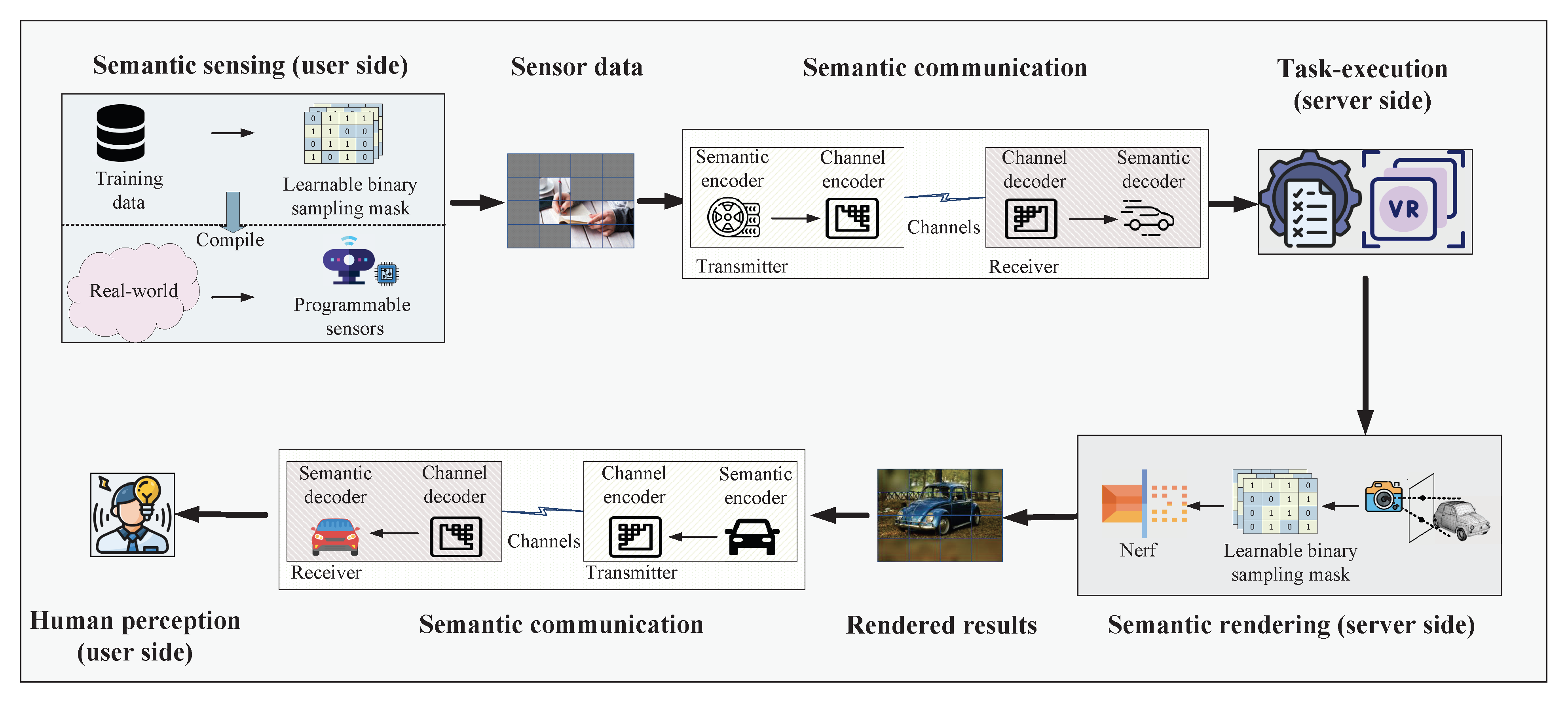}
\caption{Semantic sensing, rendering, and communication architecture}
\label{fig2}
\end{figure*}
Rendering is a time-consuming task. A fast rendering process should be designed to help the system satisfy the latency requirements. Inspired by images/videos compressed sensing~\cite{martel2020neural}, we find that existing rendering-then-compression pipeline is also sub-optimal, where each pixel in the rendered images/videos is computed with great efforts, but some pixels will later be discarded in the compression process because of inter-pixel semantic redundancy. Therefore, it will be more efficient if we can jointly consider the rendering process and the compression process. Different from the sensing process, where the real environment is unknown before sensed, the rendering process can be guided by pre-stored 3D models of objects or scenes.

\subsubsection{Design Principles}
Semantic rendering is proposed to dynamically adjust the rendering costs for each pixel in rendered images and 3D models of objects/scenes according to their semantic importance. To make it clear, we will explain the semantic rendering by the widely used volumetric neural render, NeRF~\cite{mildenhall2021nerf,xu20223d}. In NeRF, each 3D scene is first represented by a fully-connected deep network. Then, for a given viewing direction, a certain number of camera-rays will emit from a virtual camera. If the expected resolution of rendered images is $H\times W$, then there will be $H\times W$ rays in total. Each ray represents one pixel here. To get the color value of each pixel, NeRF will divide the corresponding ray into $N$ evenly spaced bins, and draw a sample uniformly from each bin. Then the deep network will generate an emited color value and a density value for each sample. The final color is calculated from the color/density values of the $N$ samples. After processing each ray, a rendered image is generated. Therefore, $H\times W\times N$ samples will be passed through the deep network during the rendering process. Considering $N=192$ used in~\cite{mildenhall2021nerf} and the spatial sizes of high-resolution images, this rendering process will cause significant latency.

In the semantic rendering, the important pixels at the semantic level will use more samples to estimate the corresponding color values. To achieve this goal, we propose to learn a 3D binary mask for the $H\times W\times N$ samples, and only the samples with +1 values in the mask will be actually used for color value calculation, as shown in Fig.~\ref{fig2}. This setting will fasten the rendering speed at the cost of a spatially-varying rendering accuracy. 
At the same time, a image restoration network will be jointly leaned with the 3D binary mask, which will recover the discarded information during the rendering process by exploring the inter-pixel semantic redundancy. The image quality after the restoration network will be used to guide the design of the 3D binary mask with the goal of maximizing the sparsity level under an image quality constraint. 

\subsubsection{Architectural Details}
To implement semantic rendering, a mask generation network will be deployed at the server, which is usually designed for each 3D model of scene/object. The best 3D binary masks for the rendering process is dependent on the viewing directions. To get the 3D binary masks for different viewing directions, the mask generation network takes the viewing directions as inputs, and outputs the mask details. As mentioned above, this mask-generation network will be trained with the goal of finding the best trade-off between reconstructed images' or 3D objects' qualities and the sparsity level of the binary masks. Similarly, the semantic rendering module should also be jointly designed with the semantic communication modules, so that the latter can deal with different rendering polices. 

\subsection{Semantic Communications}
In wireless ultimate XR, sensor data transmission and rendered results transmission pose new challenges to the existing wireless communication systems, where a great amount of data should be transmitted in a few milliseconds. To address this issue, we can use semantic communication for the transmission of the sensed data and rendered results.
\subsubsection{Design Principles}
In the uplinks of wireless ultimate XR, sensed data is usually transmitted for executing some computing tasks at the server side. Since only parts of information inside the sensed data are actually used during the task-execution process, semantic communications can be used to reduce the data traffic by extracting task-related information from the sensed data. For example, in the VR conferencing, transmitting information that represents users' body shapes, gesture, facial expressions, and dressings are enough for remote servers to update users' images in virtual conferencing room, which consumes much less wireless resource than transmitting the original camera-captured images. This idea can be extended to other computing tasks, such as hand pose estimation, light source estimation, and object detection. 

Semantic communications can also be used to improve the data transmission efficiency of the downlink connections in wireless ultimate XR. Generally speaking, the rendered images or 3D models of objects/scenes at adjacent time stamps are quite similar with minor changes caused by users' movements or virtual world changes. Therefore, instead of transmitting a new image or 3D model of object/scene, transmitting the differences between time slots is more efficient. Note that different from the video transmission methods, the differences here are the changes of semantic features \cite{jiang2022wireless}, which is much less than the original representations. Besides, in XR applications, multiple users may perceive the same virtual world from different viewing angles. In this case, the virtual contents delivered to different users are semantically-similar. By sharing the semantically-similar information on the broadcast channel \cite{9448698} and delivering semantically-different information to each user on per-user specific channel, we can reduce the overall downlink data size.  

\subsubsection{Architectural Details} 
A semantic-aware communication system is composed of a semantic encoding module, which is used to extract semantic information from the source data to transmit, and a semantic decoding module that uses the reconstructed semantic information for task execution or human perception, as shown in Fig.~\ref{fig2}. The architectural details of the semantic encoding/decoding modules usually depend on the type of source data and specific task. In particular, the semantic coding modules for image-type sensed data/rendered results transmission are usually designed based on convolutional neural networks (CNNs) or visual transformers (VTs), which have shown excellent performance in extracting semantic features from images. For the transmission of 3D models of objects/scenes, which has not attracted wide attentions, graph neural networks (GNNs) are good choices, as the points clouds of 3D objects or scenes can be represented by graphs. In addition to using GNNs, implicitly representing the 3D models of objects/scenes by a neural network~\cite{mildenhall2021nerf,xu20223d}, and semantically encoding and decoding the network parameters is another promising direction.          

Besides semantic coding modules, source coding and channel coding modules are still required to ensure efficient delivery of the semantic information over noisy wireless channels, as shown in Fig.~\ref{fig2}. The source coding module aims to find a more compact representation of the semantic information statistically. This is usually achieved through a variable-length coding mechanism, based on the criterion that a shorter code shall be assigned to semantic information more likely to appear statistically. A practical source coding method for semantic representations is the deep compression methods proposed in recent years~\cite{mentzer2020high}. 
As the interfaces between deep compression methods (served as source codes) and channel codes are digital, the channel coding module has to be designed separately from the semantic coding modules and deep compression module, resulting in a separation-based semantic communication system (SSC). 

Recently, deep joint semantic-channel coding (DeepSC)~\cite{xie2021deep} has been widely adopted in the recent semantic communication works, instead of using separated source/channel coding modules. These works use deep neural networks to represent the source coding and channel coding together without explicitly differentiating the source coding and channel coding.

After choosing the network architectures, a training loss will be used to guide the training process of the semantic-source coding networks in separation-based methods and the deep joint semantic and channel coding networks. In semantic communications, the choices of the training loss are task-dependent. 
For the sensed data transmission, task accuracy will be used as the training loss. 
As for the rendered results transmission, objective perceptual losses or subjective evaluations are exploited. Specifically, a learned perceptual image patch similarity (LPIPS) loss combined with a generative adversarial network (GAN) loss can be used as a perceptual loss of images~\cite{mentzer2020high}. GAN losses are also used in human mesh reconstruction for better visual quality~\cite{kanazawa2018end}.

\section{Case Studies}
\label{section IV}
In this section, we will demonstrate how the proposed semantic framework can be used to overcome the data rate challenges in wireless ultimate XR. In particular, we focus on the sensed data transmission in the uplink for 3D human mesh recovery task, and rendered image transmission in the downlink.

\subsection{Uplink 3D Human Mesh Transmission}
\begin{figure*}[!t]
\centering
\subfloat[Traditional communications.]{
	\label{subfig:tradition}
	\includegraphics[width=0.9\textwidth]{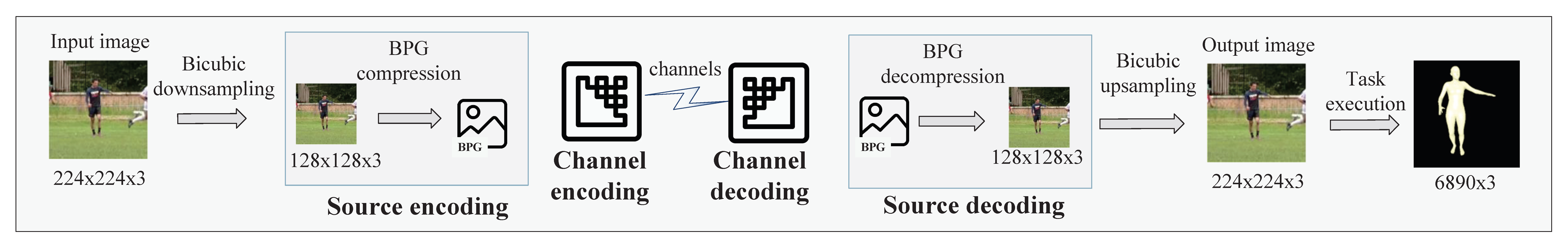} } 

\subfloat[Semantic communications.]{
	\label{subfig:semantic}
	\includegraphics[width=0.9\textwidth]{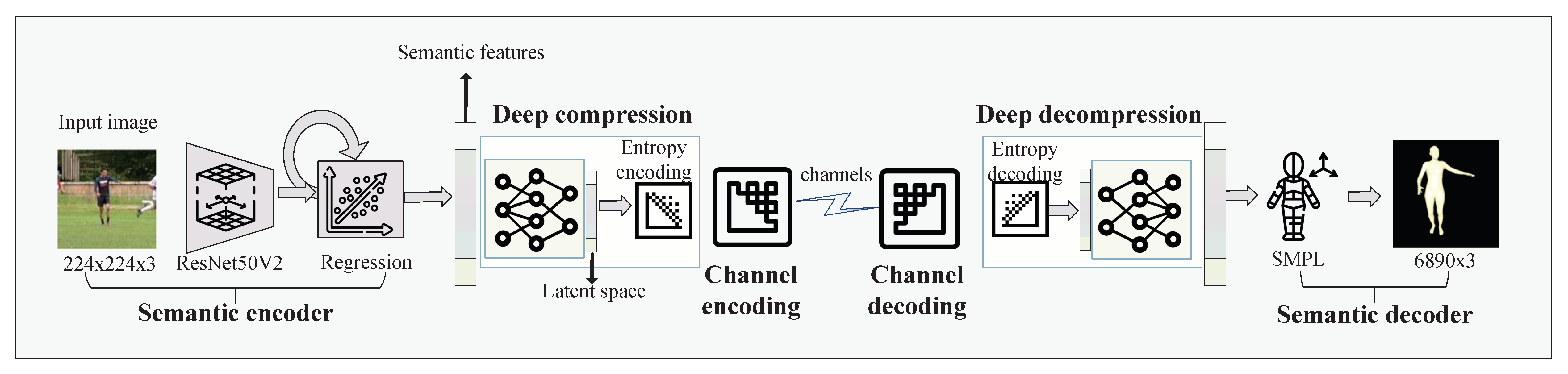} } 

\caption{Semantic communications VS traditional communications in 3D human mesh reconstruction}
\label{fig3}
\end{figure*}

\begin{figure}[!t]
\centering
\includegraphics[width=0.4\textwidth]{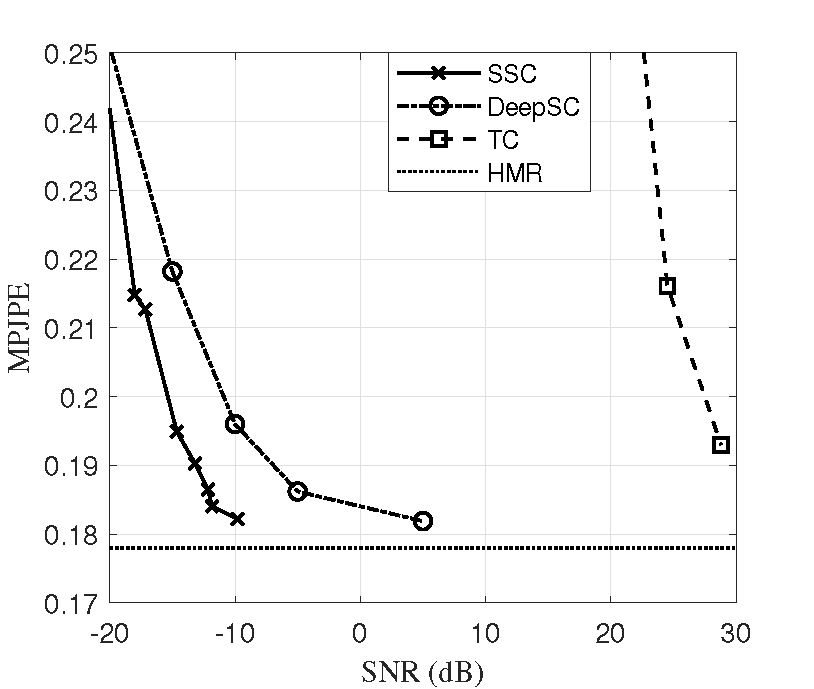}
\caption{Performance comparison between semantic communications and traditional communication on 3D human mesh construction task.}
\label{fig6}
\end{figure}
3D human mesh reconstruction is an important task in ultimate VR conferencing and gaming, where the 3D virtual images of users should be consistent with their poses and body shapes in reality. To realize the synchronous update between real images and virtual images, a camera sensor shall be deployed at the user side, keep capturing the real images of users, and transmit the camera-captured images to remote servers. The server will then estimate the user's body shape and pose from the received images, and reconstruct the 3D human mesh. In this case study, we simulate the wireless ultimate XR with the proposed framework, and compare its performance with the case with conventional communications under different channel conditions.

In the traditional communications (TC) shown in Fig.~\ref{subfig:tradition}, the color images are first bicubically-downsampled, and then compressed using image compression method BPG. The compressed data file is then transmitted through a capacity-achieving channel code \cite{jankowski2020wireless}. At the server side, the original RGB image is reconstructed in a reverse way. After that,
the server constructs the 3D human mesh using a 3D human mesh reconstruction network, HMR~\cite{kanazawa2018end}. 


In semantic communications, the 3D human body shape and pose information are extracted and transmitted, rather than the original color image. As shown in Fig.~\ref{subfig:semantic}, the semantic encoder consists of a ResNet50 and a regressor~\cite{kanazawa2018end}. The ResNet50 is used to extract the high-level convolutional features from the RGB input. The regressor then maps the features to a smaller feature vector of size $85$. These $85$ features are composed of camera angle information ($3$ features), body shape information ($10$ features), and body pose information ($72$ features). 

After obtaining the $85$ semantic features, we consider both the SSC and DeepSC methods for transmitting these features. In the SSC method, deep compression is used as the source coding module~\cite{mentzer2020high}. Specifically, a FCN is used to map the semantic features to a compact latent space with $10$ dimensions, and then entropy coding methods are used to encode features in the latent space. After the source coding modules, capacity-achieving channel codes are considered. At the receiver side, the $85$ semantic features are first reconstructed, which is then applied to predict the 3D human mesh through a SMPL model~\cite{kanazawa2018end}. In the DeepSC method, we use a 5-layer FCN with $1024$ hidden neurons to map the semantic features to modulated symbols, whose number is set as two times the channel bandwidth~\cite{xie2021deep}. After receiving the symbols, a 5-layer FCN is used to recover the semantic features, which are then passed into a SMPL model.


We use the MPI-INF-3DHP dataset~\cite{mehta2017monocular} and the losses developed in~\cite{kanazawa2018end} to train the neural networks used in the above methods. We set the bandwidth as $1$ kHz, and change the signal-to-noise-ratio (SNR) of transmitted symbols over additive white Gaussian noise (AWGN) channels from -20dB to 30dB. We use mean per joint position error (MPJPE) as the evaluation metric, where a smaller MPJPE value represents a better performance. We compare their performance in Fig~\ref{fig6}, where HMR represents the performance of human mesh reconstruction network developped in~\cite{kanazawa2018end} when there is no noise. From Fig~\ref{fig6}, semantic communication systems work better than the traditional method, denoting the efficiency of the proposed framework.

\subsection{Downlink VR Content Delivery: Broadcast}
\begin{figure}[!t]
\centering
\includegraphics[width=0.45\textwidth]{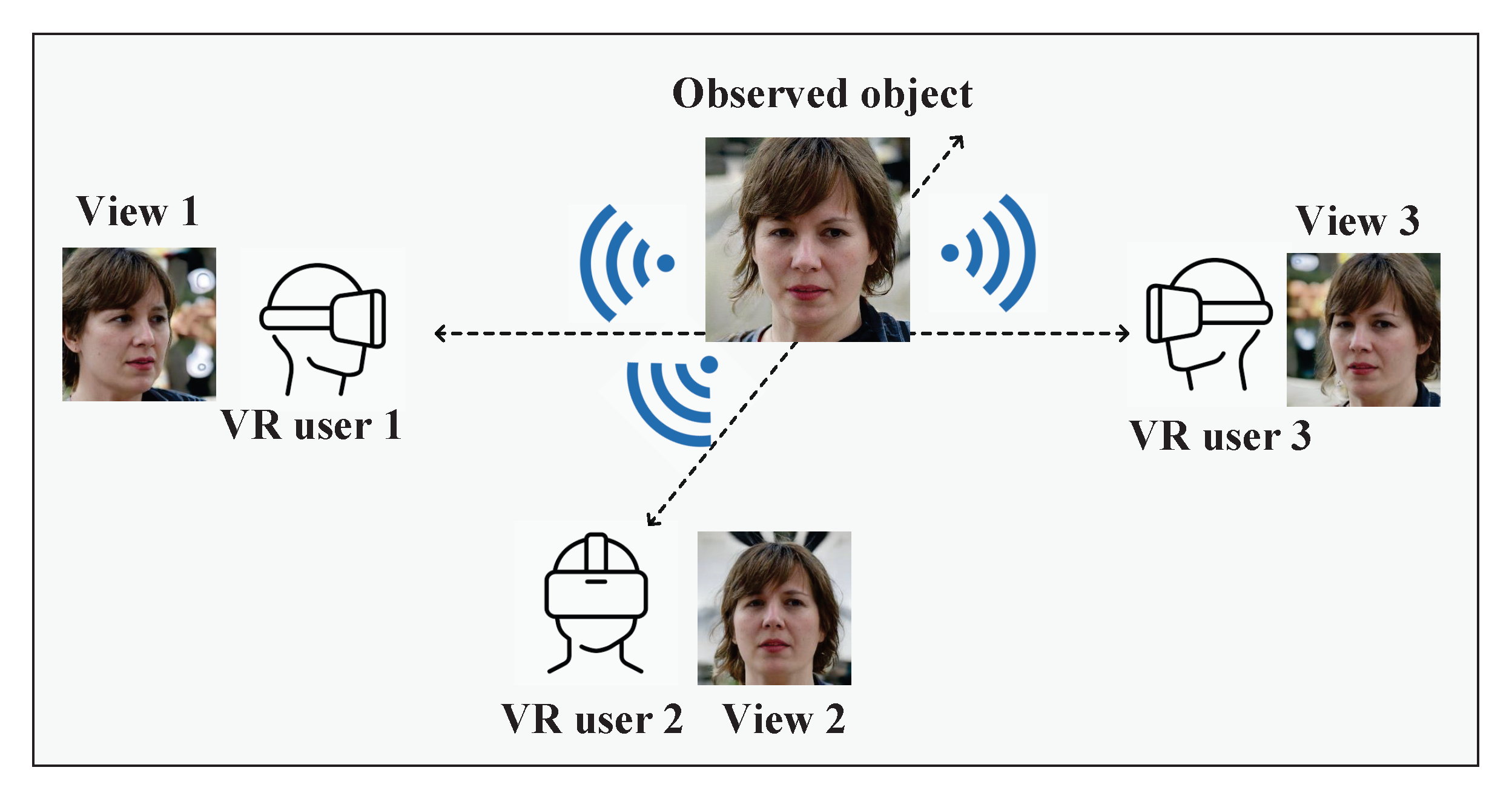}
\caption{In a wireless VR scheme, the content provider broadcasts the single-view input to the users; The users apply a 3D-aware generative model to synthesize required VR content by viewports.}
\label{fig4}
\end{figure}
In this case study, we use broadcasting transmission method alongside with the 3D-aware view synthesizing techniques \cite{xu20223d} to reduce the transmission data rates. Specifically, multi-users may watch the same VR contents from different places in the virtual world. For example, many attendees at different locations may watch someone presenting the works in VR conferencing. The 3D models of the room and the attendees are fixed during this presentation period. Therefore, instead of rendering and transmitting a specific view for each attendee, we semantically encode the 3D scene, broadcast the semantic representations of the scene, and allow users to synthesize their views locally using the broadcast information. In this way, multiple users can be served simultaneously with the shared wireless resource.  

We consider the situations where multiple users sit/stand closely in the virtual world and have very close viewing directions \cite{9448698}. In this case, it is possible to synthesize multiple views by transmitting a single view of the scene. As shown in Fig. \ref{fig4}, the VR content provider broadcasts a single-view image to a group of users. The developed 3D-aware synthesizing model enables the group to synthesize the required content based on their own viewpoints. 

Considering the increasing privacy concern and local device computing resources, a federated learning algorithm is proposed to enable the practical training of the view-synthesizing model \cite{9448698}. We also divide the models into two parts and each client is only responsible for the model update of one part. To be more exact, the volumetric rendering and color prediction parts are updated by some clients. The viewpoint analysis and camera control parts of the model are updated by the others. With this task division, the computing cost of each user can be significantly reduced.

\section{Conclusions and Future Directions}
\label{IIII}
In this article, we developed the framework with semantic sensing, rendering, and communication to support wireless ultimate XR. To bring the semantic-aware wireless ultimate XR from theory to practice, the 
following challenges should be addressed:

 \textbf{1) Trade-off between performance and computing costs:} Existing semantic coding modules have shown great success in data size reduction. However, running semantic coding modules on HMDs usually takes a long time due to the limited on-board computing resources. To reduce the running time, one potential way is to consider model distillation to find the best trade-off between the model performance and computing costs.    
 
 \textbf{2) Task-specific network generation:} Most of the existing semantic sensing or communication modules are designed in a task-specific way. This will result in a great amount of semantic models to be stored at HMDs, which usually have limited storage space. One potential way to reduce the memory costs is to use meta-learning to train a network-generation network, which dynamically generates the optimal network architectures and parameters for each task in wireless ultimate XR.  

 \textbf{3) Semantic-aware resource allocation:} In a practical communication systems, users of a particular XR application need to share wireless resource with others. The data generated by each XR user also has various size and different semantic-importance level. Therefore, it is challenging to find the optimal resource allocation policy in this complicated situation. To maximize the resource utilization efficiency, deep reinforcement learning could be considered.   

\bibliographystyle{IEEEtran}
\bibliography{ref}

\begin{thebibliography}{10}
\providecommand{\url}[1]{#1}
\csname url@samestyle\endcsname
\providecommand{\newblock}{\relax}
\providecommand{\bibinfo}[2]{#2}
\providecommand{\BIBentrySTDinterwordspacing}{\spaceskip=0pt\relax}
\providecommand{\BIBentryALTinterwordstretchfactor}{4}
\providecommand{\BIBentryALTinterwordspacing}{\spaceskip=\fontdimen2\font plus
\BIBentryALTinterwordstretchfactor\fontdimen3\font minus
  \fontdimen4\font\relax}
\providecommand{\BIBforeignlanguage}[2]{{%
\expandafter\ifx\csname l@#1\endcsname\relax
\typeout{** WARNING: IEEEtran.bst: No hyphenation pattern has been}%
\typeout{** loaded for the language `#1'. Using the pattern for}%
\typeout{** the default language instead.}%
\else
\language=\csname l@#1\endcsname
\fi
#2}}
\providecommand{\BIBdecl}{\relax}
\BIBdecl

\bibitem{akyildiz2022wireless}
I.~F. Akyildiz and H.~Guo, ``Wireless extended reality ({XR}): Challenges and
  new research directions,'' \emph{ITU J. Future Evol. Technol}, vol.~3, 2022.

\bibitem{morin2022toward}
D.~G. Mor{\'\i}n, P.~P{\'e}rez, and A.~G. Armada, ``Toward the distributed
  implementation of immersive augmented reality architectures on {5G}
  networks,'' \emph{IEEE Commun. Mag.}, vol.~60, no.~2, pp. 46--52, 2022.

\bibitem{qin2021semantic}
Z.~Qin, X.~Tao, J.~Lu, W.~Tong, and G.~Y. Li, ``Semantic communications:
  Principles and challenges,'' \emph{arXiv preprint arXiv:2201.01389}, 2021.

\bibitem{jankowski2020wireless}
M.~Jankowski, D.~G{\"u}nd{\"u}z, and K.~Mikolajczyk, ``Wireless image retrieval
  at the edge,'' \emph{IEEE J. Sel. Areas Commun.}, vol.~39, no.~1, pp.
  89--100, 2020.

\bibitem{jiang2022wireless}
P.~Jiang, C.-K. Wen, S.~Jin, and G.~Y. Li, ``Wireless semantic communications
  for video conferencing,'' \emph{arXiv preprint arXiv:2204.07790}, 2022.

\bibitem{li2021lighting}
J.~Li, H.~Li, and Y.~Matsushita, ``Lighting, reflectance and geometry
  estimation from 360 panoramic stereo,'' in \emph{Proc. IEEE Int. Conf.
  Comput. Vis. Pattern Recognit. (CVPR)}, 2021, pp. 10\,586--10\,595.

\bibitem{martel2020neural}
J.~N. Martel, L.~K. Mueller, S.~J. Carey, P.~Dudek, and G.~Wetzstein, ``Neural
  sensors: Learning pixel exposures for hdr imaging and video compressive
  sensing with programmable sensors,'' \emph{IEEE Trans. Pattern Anal. Mach.
  Intell.}, vol.~42, no.~7, pp. 1642--1653, 2020.

\bibitem{yang2018admm}
Y.~Yang, J.~Sun, H.~Li, and Z.~Xu, ``{ADMM-CSNet}: A deep learning approach for
  image compressive sensing,'' \emph{IEEE Trans. Pattern Anal. Mach. Intell.},
  vol.~42, no.~3, pp. 521--538, 2018.

\bibitem{mildenhall2021nerf}
B.~Mildenhall, P.~P. Srinivasan, M.~Tancik, J.~T. Barron, R.~Ramamoorthi, and
  R.~Ng, ``{N}e{RF}: Representing scenes as neural radiance fields for view
  synthesis,'' \emph{Commun. ACM}, vol.~65, no.~1, pp. 99--106, 2021.

\bibitem{xu20223d}
Y.~Xu, S.~Peng, C.~Yang, Y.~Shen, and B.~Zhou, ``3d-aware image synthesis via
  learning structural and textural representations,'' in \emph{Proc. IEEE Int.
  Conf. Comput. Vis. Pattern Recognit. (CVPR)}, 2022, pp. 18\,430--18\,439.

\bibitem{9448698}
Y.~Guo and Z.~Qin, ``{Federated learning for multi-view synthesizing in
  wireless virtual reality networks},'' in \emph{Proc. IEEE 96th Veh. Technol.
  Conf. (VTC Fall)}, 2022, to appear.

\bibitem{mentzer2020high}
F.~Mentzer, G.~D. Toderici, M.~Tschannen, and E.~Agustsson, ``High-fidelity
  generative image compression,'' \emph{Advances Neural Info. Process. Systems
  (NIPS)}, vol.~33, pp. 11\,913--11\,924, 2020.

\bibitem{xie2021deep}
H.~Xie, Z.~Qin, G.~Y. Li, and B.-H. Juang, ``Deep learning enabled semantic
  communication systems,'' \emph{IEEE Trans. Signal Process.}, vol.~69, pp.
  2663--2675, 2021.

\bibitem{kanazawa2018end}
A.~Kanazawa, M.~J. Black, D.~W. Jacobs, and J.~Malik, ``End-to-end recovery of
  human shape and pose,'' in \emph{Proc. IEEE Int. Conf. Comput. Vis. Pattern
  Recognit. (CVPR)}, 2018, pp. 7122--7131.

\bibitem{mehta2017monocular}
D.~Mehta, H.~Rhodin, D.~Casas, P.~Fua, O.~Sotnychenko, W.~Xu, and C.~Theobalt,
  ``Monocular 3d human pose estimation in the wild using improved cnn
  supervision,'' in \emph{Int. Conf. 3D Vis. (3DV)}, 2017, pp. 506--516.

\end{thebibliography}
\small
\end{document}